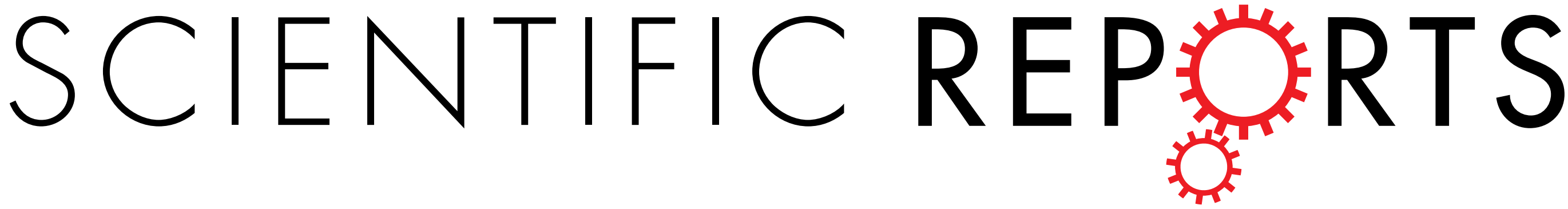

OPEN

# Real-time quasi-distributed fiber optic sensor based on resonance frequency mapping

Gyeong Hun Kim[1], Sang Min Park[1], Chang Hyun Park[1], Hansol Jang[1], Chang-Seok Kim[1] & Hwi Don Lee[2]

**Distributed optical fiber sensors (DOFS) based on Raman, Brillouin, and Rayleigh scattering have recently attracted considerable attention for various sensing applications, especially large-scale monitoring, owing to their capacity for measuring strain or temperature distributions. However, ultraweak backscatter signals within optical fibers constitute an inevitable problem for DOFS, thereby increasing the burden on the entire system in terms of limited spatial resolution, low measurement speed, high system complexity, or high cost. We propose a novel resonance frequency mapping for a real-time quasi-distributed fiber optic sensor based on identical weak fiber Bragg gratings (FBG), which has stronger reflection signals and high sensitivity to multiple sensing parameters. The resonance configuration, which amplifies optical signals during multiple round-trip propagations, can simply and efficiently address the intrinsic problems in conventional single round-trip measurements for identical weak FBG sensors, such as crosstalk and optical power depletion. Moreover, it is technically feasible to perform individual measurements for a large number of quasi-distributed identical weak FBGs with relatively high signal-to-noise ratio (SNR), low crosstalk, and low optical power depletion. By mapping the resonance frequency spectrum, the dynamic response of each identical weak FBG is rapidly acquired in the order of kilohertz, and direct interrogation in real time is possible without time-consuming computation, such as fast Fourier transformation (FFT). This resonance frequency spectrum is obtained on the basis of an all-fiber electro-optic configuration that allows simultaneous measurement of quasi-distributed strain responses with high speed (>5 kHz), high stability (~2.4 με), and high linearity ($R^2$ = 0.9999).**

Over the last two decades, fiber optic sensors have emerged as one of the fastest growing and most researched areas among modern monitoring technologies. In particular, distributed fiber optic sensing techniques, based on Raman[1], Brillouin[2,3], and Rayleigh[4] scattering within the optical fibers, have been successfully adopted in a wide range of strain- or temperature-sensing applications owing to their advantages of a large number of sensing points and a long sensing range[5]. However, ultraweak scattering signals constitute an inevitable problem that increases the burden on various sensing systems in terms of low measurement speed (below a few hertz) and a low spatial resolution (~1 m)[6]. To overcome such issues, novel technical solutions, such as optical correlation-domain scanning[3] and optical frequency-domain scanning based on optical interference[4], have been proposed. These techniques provide a higher spatial resolution (in the sub-millimeter range); however, they suffer from some drawbacks, such as increased cost, high system complexity, and reduced sensing range[6].

On the other hand, fiber Bragg grating (FBG) sensors have stronger reflection signals within an optical fiber, which can be used to acquire multiple physical and chemical parameters from discretized local points of a few millimeters along a single optical fiber[7–9]. In general, conventional FBG sensors can be simultaneously multiplexed at high measurement speeds (of the order of kilohertz)[10,11]. However, the FBG should be designed to guarantee a measurable wavelength range of the interrogator and non-overlapping spectra between the Bragg wavelengths in the wavelength domain[8]. The maximum available number of FBG sensors is limited to a few tens or less, which is a major drawback of FBG interrogation systems based on wavelength-division multiplexing (WDM)[10–13]. For example, it is difficult to interrogate more than 20 FBG sensors covering a maximum strain range of 5000 με with an optical light source having a bandwidth of 100 nm by the wavelength-division multiplexing method. Another critical disadvantage of conventional FBG sensors is that the writing of periodic grating, to ensure high reflectivity at different Bragg wavelengths, entails high labor costs, which limits mass production and cost-effective manufacturing processes[14,15].

[1]Department of Cogno-Mechatronics Engineering, Pusan National University, Busan, 46241, Korea. [2]Advanced Photonics Research Institute, Gwangju Institute of Science and Technology, Gwangju, 61005, Korea. Correspondence and requests for materials should be addressed to C.-S.K. (email: ckim@pusan.ac.kr) or H.D.L. (email: rahido@gist.ac.kr)

Recently, the limits of multiplexing capability and mass productivity have been overcome by identical-wavelength weak-reflectivity FBG array sensors[16–23]. In this array, FBGs with identical Bragg wavelengths are written with the same period grating and weak reflectivity (less than 5%); hence, serial FBGs can be continuously and economically fabricated without changing the manufacturing conditions. This fabrication process involves on-line ultraviolet (UV) exposure during the optical fiber drawing process[24–26], or femtosecond-laser writing on a single optical fiber[27–29]. Because identical FBGs can be immediately manufactured without applying conventional stripping and recoating processes to the optical fiber, the production time can be reduced, and the pristine mechanical strength of the optical fiber is maintained[14,27]. Weak FBGs have a shorter grating length and require lower UV exposure time compared to highly reflective FBGs, which is convenient for alignment and reduces the processing time[9]. In addition, with single manufacturing systems of an identical weak FBG array, the additional fiber-splicing process of individual FBGs can be avoided, which not only reduces the splicing loss significantly but also improves the sensitivity and capability of the interrogation system[25].

However, conventional WDM interrogators of FBGs cannot be adapted to identical weak FBG sensors, because all the FBGs along the optical fiber have identical Bragg wavelengths. Currently, an identical weak FBG array is typically interrogated by two types of methods: time-division multiplexing (TDM)[16–19] and optical frequency-domain reflectometry (OFDR)[20–22]. In TDM interrogations, a pulse-modulated laser[16–19] is employed to multiplex each identical FBG on the basis of a time-of-flight (TOF) method. The quantitative strain or temperature response of individual FBGs have to be measured by spectroscopic instruments, such as a tunable laser[16–18] or an optical spectrum analyzer (OSA)[19]. Thus, the measurement speed is limited to a few tens of hertz owing to the slow response time of the spectroscopic instruments and the individual scanning of all the FBGs. To overcome the drawback of low measurement speed of TDM methods, the dispersive Fourier transformation technique[30–33] is adopted to rapidly obtain strain information at up to 20 kHz[23]. However, as with other TDM methods, it is difficult to widely apply this technique to various fields because it requires high-performance instruments, such as a high-speed oscilloscope, fast and highly sensitive photodetectors, and optical amplifiers. In some studies, identical weak FBGs have been interrogated using the OFDR method, which is suitable for densely distributed measurement applications owing to its high spatial resolution (of the order of millimeters) and limited sensing length (of the order of meters)[20–22]. The maximum sensing length and measurement speed are determined by the coherence length and sweep rate of the tunable laser source[20–22]. Reflection distribution along an optical fiber requires high-performance data computing capabilities, which reduce the feasibility of real-time measurement. In addition, such interrogation methods based on single round-trip measurement for the identical weak FBG array suffer from the intrinsic crosstalk problem owing to multiple reflections between each identical FBG[21]. Hence, the identical FBGs in the sensing array must have ultra-low reflectance (less than 0.01%), which severely degrades the SNR[18]. Moreover, when an optical signal propagates across an identical FBG array, the signal power decreases exponentially[17]. This problem is especially severe in the case of FBGs located backward in an array of identical FBGs.

In this study, we demonstrate a real-time quasi-distributed fiber optic sensor system based on resonance frequency mapping for simultaneous multiplexing and strain measurement of an identical weak FBG array. Resonance configurations based on a ring cavity can simply and efficiently address the intrinsic problems of identical weak FBG sensors, such as crosstalk and optical power depletion, thereby reducing the cost and complexity of conventional single round-trip measurement systems. When a modulation frequency fed into the gain of the laser cavity is synchronized with the resonance frequency of the cavity formed by each individual FBG in the sensing array, a synchronized FBG can be selected and isolated from the other FBGs. Thus, identical weak FBGs can be separately interrogated with relatively high SNR, low crosstalk, and low optical power depletion. In addition, this technique is based on an active mode-locked fiber laser cavity, enclosing a chirped FBG (CFBG) for chromatic dispersion. Instead of detecting the reflection spectra of FBGs in the wavelength domain, or the TOF of FBGs in the time domain, this method multiplexes FBGs on the basis of the resonance frequencies induced by the total cavity lengths determined by the positions of the FBGs. The highly dispersive CFBG induces a significant change in the total cavity length when a strain or temperature variation is applied to each FBG. Therefore, strain or temperature variation of the FBG can be measured by detecting the shift in the resonance frequency resulting from the change in the Bragg wavelength. The proposed quasi-distributed fiber optic sensor based on resonance frequency mapping is not only more compact and cost-effective but also delivers better performance than other identical weak FBG interrogation systems. The improved sensing performance of the identical weak FBG array was demonstrated experimentally. The results confirmed the large capacity (greater than 31 for identical FBGs and up to 770), high sampling rate (greater than 5 kHz), high SNR (higher than 25 dB), and high linearity ($R^2 = 0.9999$) of the proposed interrogator configuration.

## Principle

The principle of identical weak FBG interrogation based on resonance frequency mapping is shown in Fig. 1. The configuration of the resonance frequency mapping system consists of an active mode-locked fiber laser cavity with multiple laser cavities, employing an optical gain where the driving current is externally modulated at a modulation frequency ($f_m$) between a sensing head and a CFBG. Each identical weak FBG on the sensing head has a different resonance frequency induced by the total cavity length, which is determined by its position along the sensing head. The initial total cavity length of the $i$th FBG is $L_i$, with a Bragg wavelength of $\lambda_A$, which is equal to twice the sum of the main cavity length ($L_M$), the cavity length on the CFBG ($L_C(\lambda_A)$), and the cavity length on the sensing head of the $i$th FBG ($L_{FBG}(i)$):

$$L_i = 2 \cdot (L_M + L_c(\lambda_A) + L_{FBG}(i)) \tag{1}$$

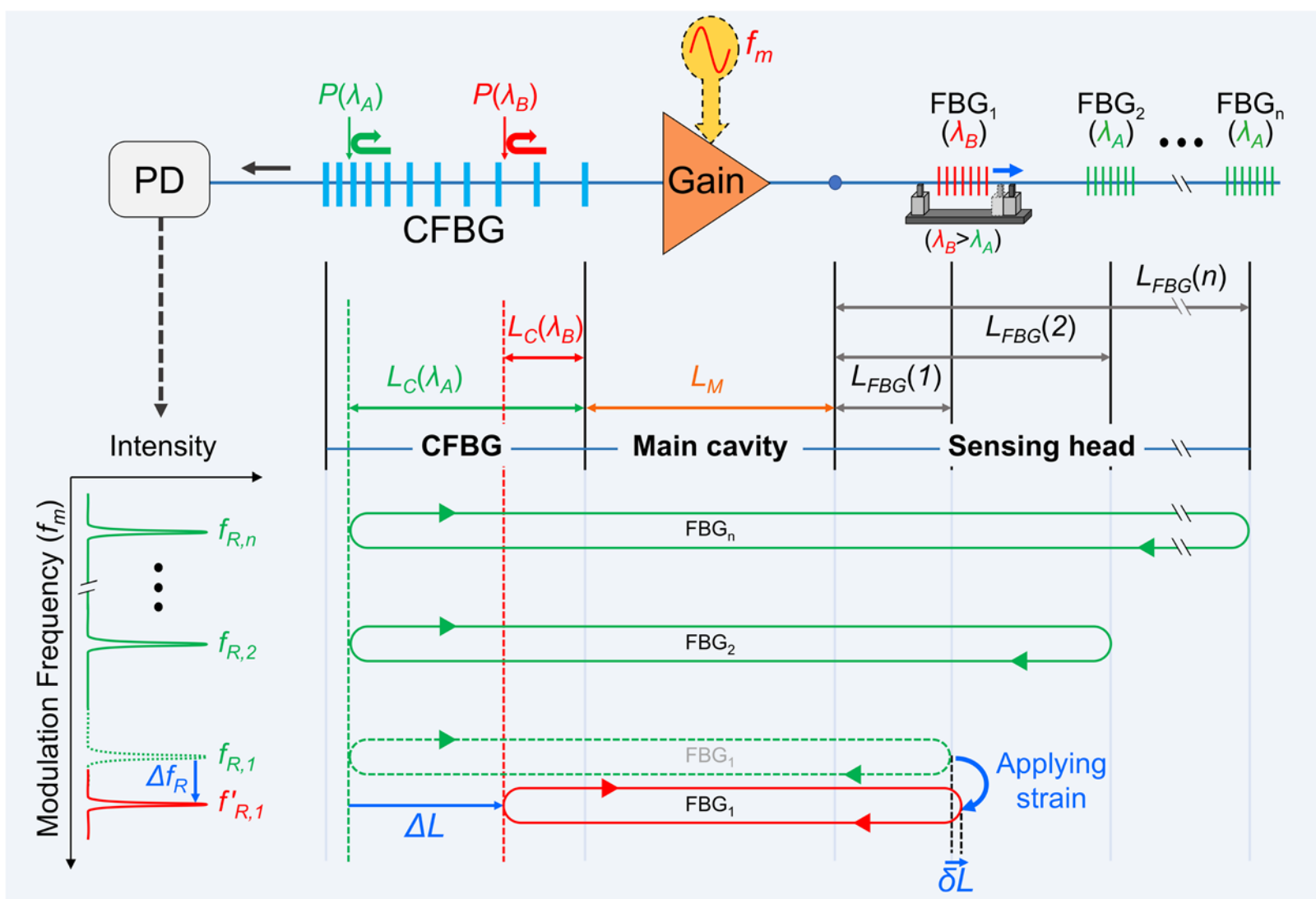

**Figure 1.** Principle of identical weak FBG interrogation based on resonance frequency mapping. Each identical weak FBG has a different resonance frequency induced by a total cavity length determined by its position along the sensing head. Strain and temperature variation of the FBGs can be measured by the shift in the resonance frequencies resulting from the change in the total cavity length ($2\Delta L$), mainly because of the relation between the change in the Bragg wavelength and the high dispersion of the CFBG.

The modulation frequency ($f_m$, repetition rate of the pulse signal) was continuously swept through the programmed frequency range. During the sweeping of the modulation frequency from the lowest to the highest values, strong output lasing powers (peak signals) were detected at each specific resonance frequency ($f_{R,i}$). This frequency was a positive-integer multiple (of the order $N$ of the resonance) of the free spectral range ($FSR$, defining the reciprocal of the round-trip time) of the laser cavity configured by each FBG. When multiplexing the $i$th FBG ($FBG_i$), $FSR_i$ and $f_{R,i}$ are related as follows:

$$FSR_i = \frac{c}{n_e \cdot L_i} \tag{2}$$

$$f_{R,i} = N \cdot FSR_i \tag{3}$$

where $L_i$ is the initial total cavity length configured by the $i$th FBG, $c$ is the speed of light in vacuum, and $n_e$ is the effective refractive index of the optical fiber.

Under steady-state conditions without strain, the initial total cavity length ($L_i$) can be accurately determined from the cavity length on the sensing head of $FBG_i$ ($L_{FBG}(i)$) and the cavity length on the CFBG ($L_C(\lambda_A)$) related to the reflection point, $P(\lambda_A)$ (note that $\lambda_A$ is the initial Bragg wavelength under steady-state conditions without strain; see Fig. 1). Therefore, the resonance frequency $f_{R,i}$ of $FBG_i$ is directly related to the initial total cavity length ($L_i$), which simultaneously includes the information of the reflection point of both $FBG_i$ and the CFBG.

Under an external strain on $FBG_i$, the change in the period of the Bragg gratings proportionally shifts the Bragg wavelength of $FBG_i$ from $\lambda_A$ to $\lambda_B$. The length of the optical cavity configured by $FBG_i$ is mainly changed by the wavelength-dependent change in the reflection point from $P(\lambda_A)$ to $P(\lambda_B)$ along the CFBG. This change in the cavity length ($2\Delta L$) results from the high chromatic dispersion of the long CFBG. The change in the cavity length on the CFBG results in a shift in the corresponding resonance frequency from $f_{R,i}$ to $f'_{R,i}$. For example, when a positive strain of 1000 $\mu\varepsilon$ with a center wavelength of 1560.3 nm and a strain-holder length of 100 mm is applied to $FBG_i$, the positional change in the reflection center is 0.05 mm ($\delta L$) and the total cavity length increases by 0.1 mm. However, the Bragg wavelength shift from $\lambda_A = 1560.3$ nm to $\lambda_B = 1561.52$ nm induces a much larger shift in the reflection point from $P(\lambda_A)$ to $P(\lambda_B)$ inside the CFBG ($\Delta L = 267$ mm), which decreases the total cavity length to 534 mm. Therefore, the change in the total cavity length, $2\Delta L$, induced by the physical strain of $FBG_i$, is negligibly smaller than that induced by the shift in the reflection point on the CFBG, $2\Delta L$ (see Supplementary Fig. 1).

The resonance frequency shift ($\Delta f_{R,i}$) is related to the strain ($\varepsilon$) on the $FBG_i$ as

**Figure 2.** Schematic of real-time identical weak FBG interrogation based on resonance frequency mapping. The optical laser cavity consists of two different semiconductor optical amplifiers (SOA), an isolator (ISO), a 10:90 optical coupler (OC), two optical circulators (CIR), a CFBG, and an array of 31 identical weak FBGs as a sensing head. The switched SOA was modulated by a short-pulse (~3 ns) driving signal by a frequency-swept pulse pattern generator (FS-PPG). The resonance frequency spectrum was repeatedly obtained as the frequency of the gain modulation was linearly swept in time.

$$\Delta f_{R,i} = N \cdot (FSR'_i - FSR_i) = N \cdot \frac{c}{n_e} \cdot \left( \frac{1}{L_i + 2\delta L + 2\Delta L} - \frac{1}{L_i} \right) \quad (4)$$

where $FSR_i$ is the FSR of the initial laser cavity, $FSR'_i$ is the FSR of the laser cavity when a strain is applied to FBGi, $\delta L$ is the shift in the reflection point of $FBG_i$ after applying the strain to $FBG_i$, and $\Delta L$ is the shift in the reflection point in the CFBG, induced by the wavelength-dependent change in the reflection point in the CFBG, which results from the Bragg wavelength shift of $FBG_i$.

By assuming that $2\Delta L \ll L_i$ and $\delta L \ll \Delta L$, Eq. (4) can be approximated as

$$\Delta f_{R,i} \approx N \cdot \frac{c}{n_e \cdot L_i^2} \cdot (-2\Delta L) = N \cdot FSR_i^2 \cdot D \cdot S \cdot \varepsilon \quad (5)$$

where $N$ is the order of the resonance, $D$ is the chromatic dispersion of the CFBG, $S$ is the strain response constant (1.2168 nm·mε$^{-1}$ at 1560 nm)[8], and $\varepsilon$ is the strain applied to the $FBG_i$.

## Results

**Experimental setup.** The experimental setup of a real-time quasi-distributed fiber optic sensor system based on resonance frequency mapping for multiplexing and measuring an identical weak FBG array is shown in Fig. 2. The main component of the interrogation system is an all-fiber ring cavity laser consisting of two different semiconductor optical amplifiers (SOA), an optical isolator (ISO), two optical circulators (CIR), 31 identical weak FBGs for the sensing head, and a 10:90 optical coupler (OC) for the laser output coupling. All the optical fiber components have low polarization sensitivity (less than 1.5 dB), ensuring system stability under arbitrary polarization perturbations. One of the SOAs amplifies the optical signals in the laser cavity, while the other modulates the optical intensity by a driving signal applied electrically. The switch SOA was specifically designed for high-speed switching (faster than 500 ps) and a high extinction ratio (greater than 40 dB). The in-line SOA was inserted to increase the low net gain of the cavity owing to the high loss and low reflectivity from the sensing head. The ISO was used to prevent the reduction in gain and the noise figure degradation of the SOAs owing to the interference between the two SOAs. The CFBG is a highly dispersive fiber component with a negative dispersion of −2152 ps·nm$^{-1}$ and a grating region length of 10 m. A 30-m delayed optical fiber ($L_d$) was inserted into the main cavity to prevent signal overlap between the first- and second-order resonance signals. For multiplexing each identical FBG along the sensing head, the switch SOA was modulated by a short-pulse (~3 ns) driving signal by a frequency-swept pulse pattern generator (FS-PPG). The maximum current of the driving signal was 180 mA. As the frequency of the gain modulation was linearly swept in time, the output power was collected by a slow photodetector (PD) having a bandwidth of 1 MHz, which repeatedly provided the resonance frequency spectrum of the sensing head in the time domain.

**Interrogation of an identical weak FBG array based on resonance frequency mapping.** The sensing head consists of an array of 31 identical weak FBGs spaced at approximately 1-m intervals along the optical fiber. Each FBG has a 3-dB bandwidth of ~0.2 nm, reflectivity of ~4%, and a Bragg wavelength of 1560.3 nm. Figure 3a shows the resonance frequency spectrum of the sensing head based on resonance frequency mapping. The main peaks of the 31 identical weak FBGs are clearly obtained at each resonance frequency, indicating that the multi-reflection crosstalk effect is significantly suppressed even though FBGs with much higher reflectivities are employed compared to other identical weak FBG interrogations[16–24]. The measured average power was not constantly maintained along the modulation frequency, because the repetition rate of the output pulse varied during the average power measurements using a low-speed PD. Therefore, we defined the power density as the calculated intensity of a single pulse, obtained by dividing the measured output power by its modulation frequency. As the modulation frequency fed into the switch SOA was linearly swept from 1.182 MHz to 1.864 MHz in a series of time cycles, the resonance frequency mapping technique successfully identified the 31 identical weak FBGs along the sensing head. From Eq. (2), it is found that the peak resonance frequency of 1.839 MHz corresponds to a total cavity length of 111.11 m. This value corresponds to the sum of the main cavity length (103.74 m), the cavity length on the CFBG (5.49 m), and the cavity length on the sensing head of $FBG_1$ (1.88 m). Similarly, the peak signal at the resonance frequency of 1.192 MHz corresponds to a total cavity length of 171.42 m for $FBG_{31}$. The SNR at the resonance frequency of $FBG_1$ (~25 dB) can be derived from the ratio of the signal intensity above the noise floor ($I_{signal}=63.57$) to the standard deviation of the noise floor ($\sigma_{noise}=0.2$). The modulation frequency (x-axis in Fig. 3a) was converted into the total cavity length using Eq. (2). As shown in Fig. 3b, the result multiplexes each FBG as position information along the optical fiber. The measured difference in the cavity lengths between adjacent FBG pairs was approximately 2 m, which is consistent with doubling the designed 1-m intervals of the adjacent FBGs along the optical fiber. Figure 3c shows the measured resonance frequencies and standard deviations of the 31 identical weak FBGs along the sensing head. As the modulation frequency is inversely proportional to the total cavity length (according to Eq. (2)), the number of optical power measurements decreases as the total cavity length increases. For identifying the resonance frequency from the Gaussian fitting of the peak signal, the standard deviation increases with decreasing modulation frequency owing to the relatively low data density. As shown in Fig. 3d, the resonance configuration effectively compensated the optical loss, which resulted from multiple reflections among the identical FBGs. Although the interrogated signal power decreased as the number of FBGs along the optical fiber increased, the slope of the measured signal depletion (−0.033 dB/number of FBGs) was 10 times smaller than that of the conventional single round-trip measurement configuration. In conventional single round-trip measurements, such as the TDM or OFDR methods[16–22], the theoretical slope of the signal depletion is −0.355 dB/number of FBGs. The ratio $I_{signal}/I_{noise}$ of the signal intensity above the noise floor ($I_{signal}$) of $FBG_1$ to the intensity of the noise floor ($I_{noise} = 0.2$) was 2.91 dB.

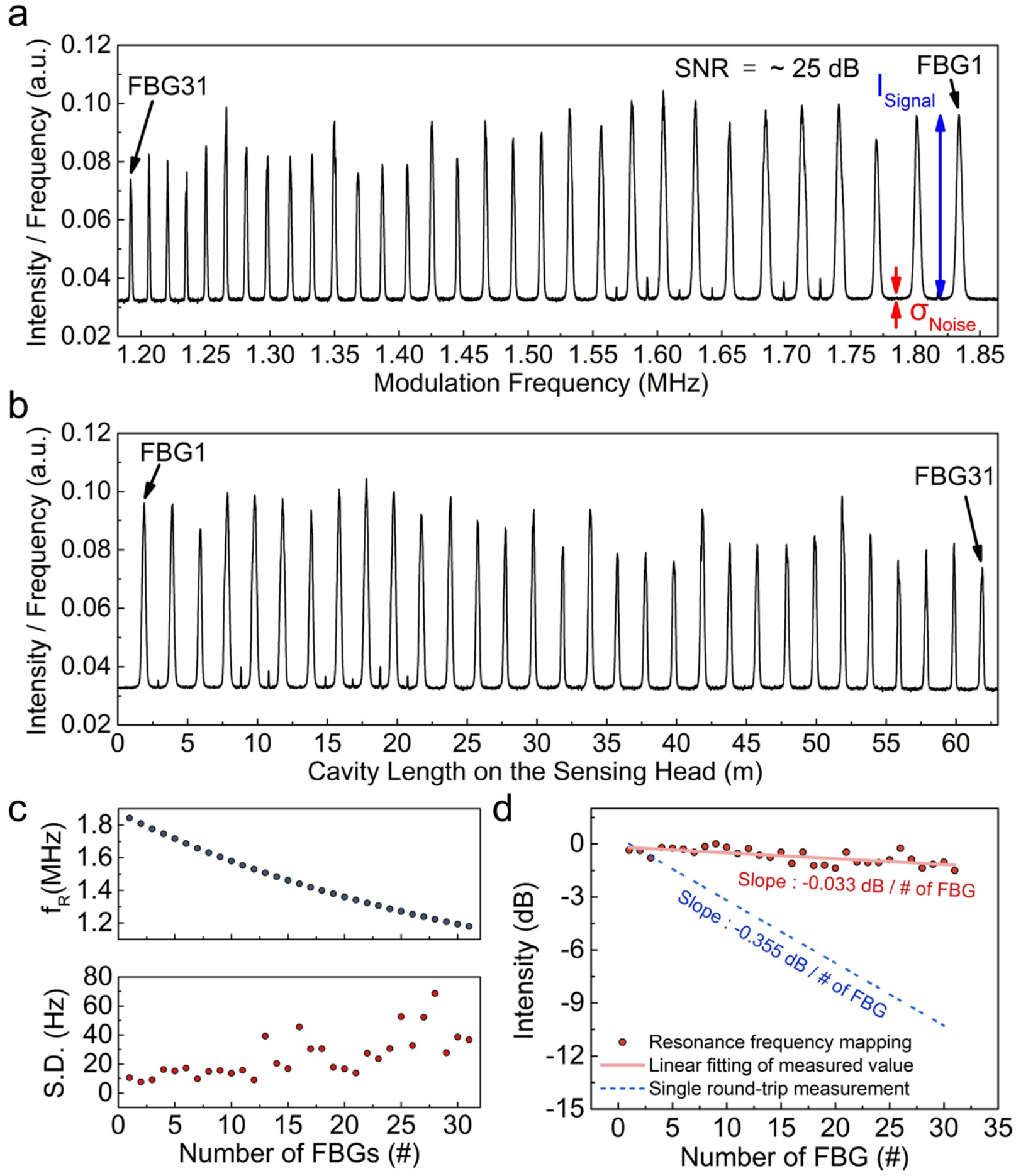

**Figure 3.** Interrogation of an identical weak FBG array: (**a**) resonance frequency spectrum of the sensing head with 31 identical weak FBGs during the sweeping of the modulation frequency across the first-order ($N=1$) resonance frequency of each FBG; (**b**) length-domain sampling graph of (a), obtained by Eq. (2); (**c**) measured resonance frequency and standard deviation of each identical weak FBG; and (**d**) comparison of the interrogation-signal power of an identical weak FBG array implemented by the single round-trip measurement and the resonance frequency mapping.

**Figure 4.** Static strain response of the resonance frequency: (**a**) resonance frequency spectra and (**b**) strain response of resonance frequency of $FBG_1$ to the applied strain increasing from 0 to 2250 με; (**c**) stability of the resonance frequency mapping system using $FBG_1$ without strain, measured for 21 min. The standard deviation of the measured strain is ~1.2 με (inset).

**Static strain response of resonance frequency.** Figure 4a,b show that the applied strain response as positive strain on $FBG_1$ increased from 0 to 2250 με (9 steps; see Supplementary Fig. 2). The initial resonance frequency, $f_{R,1}$, without strain on $FBG_1$ was measured at 1.839 MHz. As the static strain on $FBG_1$ was incremented in steps of 250 με, the resonance frequency increased to $f'_{R,1}$ with a sensitivity of 9.037 kHz·mε$^{-1}$, and the response was strongly linear ($R^2 = 0.9999$). This result is in good agreement with the theoretical sensitivity, calculated as 8.856 kHz·mε$^{-1}$ by Eq. (5) (difference = 2%, see Supplementary note 2). The dynamic range (*D.R.*) of the available strain measurements can be obtained as

$$D.\ R. = \pm \frac{n_e \cdot L_{inter}}{c \cdot D \cdot S} \tag{6}$$

where $L_{inter}$ is the distance between adjacent FBGs.

The dynamic range of the resonance frequency mapping system is proportional to the distance between adjacent FBGs ($L_{inter}$), and it is inversely proportional to the total dispersion. By setting the experimental conditions as $L_{inter} = \sim 1$ m, $D = -2152$ ps·nm$^{-1}$, and $S = 1.2168$ nm·mε$^{-1}$ in Eq. (6), the value of the *D.R.* in the present experiments was estimated as ± 1868 με.

To verify the stability of the resonance frequency mapping system, we measured the strain response of $FBG_1$ under the strain-free condition for 20 min. The strain response was obtained by converting the measured resonance frequency using the measured sensitivity (see Fig. 4c). The inset of Fig. 4c shows the probability distribution of the strain response. This plot confirms the reliable stability of the resonance frequency mapping system ($2\sigma = 2.4$ με).

**Real-time dynamic strain response of resonance frequency.** Figure 5a,b show the experimental scheme for demonstrating the dynamic strain response of the proposed resonance frequency mapping system. Four identical weak FBGs were attached to the centers of the third, fourth, fifth, and sixth strings of a guitar. The four strings (strings 6–3) were vibrated by manually plucking them in sequence at approximately 2-s intervals. The measured temporal strain variations of the sensing FBGs are shown in Fig. 5c. These plots illustrate the vibration characteristics (frequency and amplitude) of each guitar string. Figure 5d–g show the fast Fourier transformation (FFT) spectra of the measured temporal vibrations of strings 6–3. On an ideally tuned guitar, the fundamental frequencies of the open strings are uniquely determined and increase in the following order: string 6 (E = 82 Hz), string 5 (A = 110 Hz), string 4 (D = 147 Hz), and string 3 (G = 196 Hz). In the frequency analysis, the vibration frequencies including fundamental frequency (first order) and harmonic frequencies of guitar strings were successfully measured by the FBGs placed on each string. The results closely matched the ideal fundamental frequencies of 81.71 Hz, 109.45 Hz, 145.27 Hz, and 194.17 Hz for the sixth, fifth, fourth, and third strings, respectively (with an error of ~1%). For comparison, the sounds of the vibrated guitar strings were recorded by the microphone of a mobile phone, and similar frequency spectra were obtained (see Supplementary Fig. 3). Moreover, when the temporal strain variation data of each guitar string measured by the identical weak FBGs were converted into sound signals, they were clearly demodulated to the original vibrational sounds (see Supplementary Movie 1).

## Discussions and Conclusions

We proposed a real-time quasi-distributed fiber optic sensor with identical weak FBGs, based on resonance frequency mapping. The proposed system can successfully and efficiently interrogate up to 31 identical weak FBGs by detecting the resonance frequencies of the FBGs. When strain is applied to the FBGs, it affects the movement of the reflection point in the CFBG, thereby changing the specific total cavity length of the strained FBGs. By mapping the resonance frequency spectrum, the dynamic response of each identical weak FBG is rapidly acquired in the order of kilohertz, and it is directly interrogated in real time without time-consuming computation, such as fast Fourier transformation (FFT). The resonance frequency mapping approach offers several advantages. (1) It improves the cost efficiency and compactness compared to other time-domain interrogation systems with identical weak FBGs[16–19]. (2) It realizes a long-lifetime, low-maintenance system, most of whose simple units (optical fibers and semiconductor devices) are fundamental components in the optical networking industry, with typical lifetimes exceeding $5 \times 10^5$ h. (3) It provides high linearity, high speed, and high stability against harsh environmental factors (e.g., vibration, temperature, and humidity) by a fully electro-optic mechanism without any mechanical scanning.

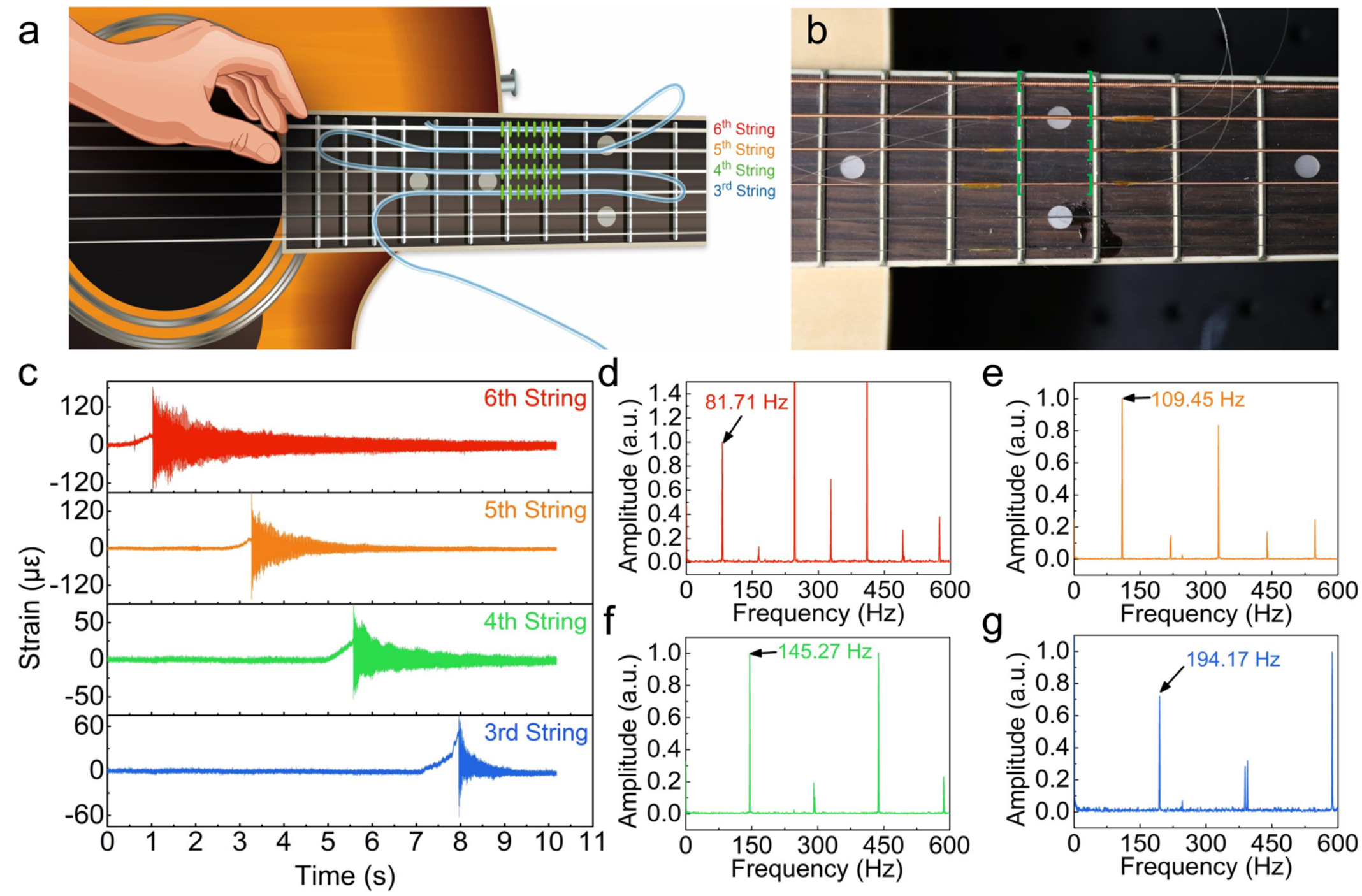

**Figure 5.** Real-time vibration detection of guitar strings: (**a**) schematic of vibration detection of guitar strings; (**b**) photograph showing the positions of the attached FBGs (green brackets); (**c**) dynamic strain responses of the attached FBGs after oscillating the open guitar strings. FFT-processed frequency spectra of the dynamic strain responses of the FBGs attached to the (**d**) sixth string (E: 82 Hz), (**e**) fifth string (A: 110 Hz), (**f**) fourth string (D: 147 Hz), and (**g**) third string (G: 196 Hz) under the ideal condition of open strings.

The multiplexing capacity in the current study is limited to several tens of identical weak FBGs, because we employed an identical weak FBG array with relatively high reflectivity in order to suppress signal depletion and crosstalk from multiple reflections among identical FBGs. The multiplexing capacity can be increased considerably (to over 700) by decreasing the reflectivity of each FBG below 0.1%, theoretically (see Supplementary Fig. 4). Moreover, when additional identical weak FBG arrays with slightly different center Bragg wavelengths are connected in series (by splicing them to the ends of the original identical weak FBG array), the multiplexing capacity can be significantly enhanced to tens of hundreds of sensing FBGs by using a single multiplexing unit (see Supplementary Fig. 5). The sweeping range of the modulation frequency can be expanded to detect the extended resonance frequencies of additional FBGs in extremely long cavities without serious technical difficulties. Moreover, to prevent the degradation of the standard deviation resulting from the decreased data density at longer total cavity lengths (see Fig. 3c), the time-linear modulation frequency sweeping can be replaced with length-linear sweeping. This modification can acquire data at equal intervals on the length axis and solve the problem of uneven data density.

In conclusion, our novel resonance frequency mapping system can simultaneously interrogate an identical weak FBG array. This technique can potentially enable us to sense various physical parameters on the basis of multiple identical FBG sensors, which is not possible with currently available distributed sensing techniques. From a mass-production perspective, this method offers two distinct advantages, i.e., a reduction in the sensor cost by employing identical weak FBGs, which are already mass-produced, and perfect simultaneous interrogation by mapping the resonance frequency spectrum. Therefore, in the future, we expect identical weak FBG sensors based on resonance frequency mapping to become more competitive and to be widely adopted in various fields.

## Methods

**Acquisition of resonance frequency spectrum of an identical weak FBG array.** The sensing head consists of 31 identical weak FBGs (SJ Photonics Inc.) written in a single-mode fiber (SMF-28, Corning) at ~1-m intervals. The reflectivity and Bragg wavelength of the FBGs were 4% and 1560.3 nm, respectively. To prevent signal overlap between the first- and second-order resonance signals ($N = 1$ and $N = 2$, respectively), we calculated the required additional main cavity length by simulating the relationship between the resonance frequency and the total cavity length (see Supplementary Fig. 6). As the length of the sensing head should be less than half of the total cavity length of $FBG_1$ ($N = 1$, see Supplementary Fig. 7 and Supplementary notes 1, 2), we added a 30-m-long delayed optical fiber (SMF-28, Corning) to the main cavity in other to increase the total cavity length of $FBG_1$ to 111.11 m, which is sufficient to multiplex the 30.15-m sensing head. As the frequency of the gain modulation was swept linearly in time across the regions of interest, the output power was collected by a slow PD (Model 2117, New Focus) with a bandwidth of 1 MHz. The sweeping rate was 100 Hz in the range of 1.1–1.9 MHz with a step frequency of 8 Hz and a frequency switch time of 100 ns. The gain of the switch SOA (IPSAD 1502, Inphenix Inc.) was modulated by an FS-PPG (TDI Inc., South Korea) and a current driver board (TDI Inc.). Short pulses (pulse width of ~3 ns) were delivered with a maximum current of ~180 mA. An in-line SOA (IPSAD 1501, Inphenix Inc.) was employed to improve the net gain of the cavity, and it compensated for the loss due to the weak reflectivity of the FBGs and multiple reflections between identical FBGs. High chromatic dispersion ($-2152\ ps\cdot nm^{-1}$ at ~1550 nm) was realized by a CFBG module (Proximion Inc.). The CFBG had a reflectivity of less than 90% and a 3-dB bandwidth of 45.5 nm. The collected signal was converted into digital signals using a digitizer (ATS9350, AlazarTech) operated at 20 $Msamples\cdot s^{-1}$.

**Static strain response measurements.** To measure the static strain response, the acrylate coating was stripped from the optical fiber of $FBG_1$, and each end of $FBG_1$ was attached to a separate micro-linear stage (M-462-X-M, Newport) using an instant adhesive (Loctite 401). The length of the fixed section was 1 m, which was sufficiently long to ensure the precision and accuracy of the strain response measurement. Static strain was applied to $FBG_1$ by moving the micro-linear stage from 0 to 2.25 mm in 250-μm steps using a high-precision motorized actuator (LTA-HS, Newport). As the strain on $FBG_1$ incrementally increased, the power density spectra were obtained by a personal computer. The resonance frequencies were identified by Gaussian-peak fitting. The fitting was performed by a built-in algorithm in the commercial software LABVIEW 2014 (National Instruments).

**Real-time dynamic strain response measurements.** The sensing head consisted of four identical weak FBGs spliced at intervals of ~0.7 m. Each FBG was fixed at the center of a different guitar string (strings six, five, four, and three). The strings were vibrated manually by plucking them in order (from the sixth to the third strings), and the resonance frequency spectra were recorded at an interrogation speed of 5 kHz over a frequency sweeping range of 6.00–7.33 MHz. The resonance frequencies of the four FBGs were also identified at each time by Gaussian-peak fitting using the built-in algorithm of LABVIEW 2014 (see Supplementary Fig. 8).

## Acknowledgements

This work was supported by the National Research Foundation of Korea (NRF) grant funded by the Ministry of Science and ICT, Republic of Korea (NRF-2015M3A9E2067143, NRF-2017R1A4A1015627), the NRF grant funded by the Ministry of Education, Republic of Korea (NRF-2018H1A2A1063133).; Fig. 5a is based on vector images obtained from freepik.com.; We thank the Taihan Fiberoptics for their technical support about the optical fiber.

## Author Contributions

The experiments were designed and performed by G.H. Kim, C.H. Park and S.M. Park. The data were analyzed and interpreted, and the figures of the results were prepared for the manuscript by G.H. Kim and H. Jang. The manuscript was written by G.H. Kim, C.-S. Kim and H.D. Lee. C.-S. Kim and H.D. Lee equally supervised the project, designed experiments, and interpreted data. All authors contributed to the critical reading of the manuscript.

## Additional Information

**Competing Interests:** The authors declare no competing interests.